\documentclass[modern]{aastex631}
\pdfoutput=1

\usepackage{amsmath,amssymb}
\begin{document}

\title{Wave Generation by Flare-Accelerated Ions and Implications for $^3$He Acceleration}

\author{A. Fitzmaurice}
\affiliation{Department of Physics, University of Maryland,
College Park, MD 20740, USA}
\affiliation{Institute for Research in Electronics and Applied Physics
University of Maryland, College Park, MD 20740, USA}

\author{J. F. Drake}
\affiliation{Institute for Research in Electronics and Applied Physics
University of Maryland, College Park, MD 20740, USA}
\affiliation{Department of Physics, the Institute for Physical Science 
and Technology and the Joint Space Institute, University of Maryland,
College Park, MD 20740, USA}

\author{M. Swisdak}
\affiliation{Department of Physics, University of Maryland,
College Park, MD 20740, USA}
\affiliation{Institute for Research in Electronics and Applied Physics
University of Maryland, College Park, MD 20740, USA}



\begin{abstract}

The waves generated by high-energy proton and alpha particles streaming from solar flares into 
regions of colder plasma are explored using particle-in-cell simulations. Initial distribution 
functions for the protons and alphas consist of two populations: an energetic, streaming 
population represented by an anisotropic ($T_{\parallel} > T_{\perp}$), one-sided kappa function 
and a cold, Maxwellian background population. The anisotropies and non-zero heat fluxes of these 
distributions destabilize oblique waves with a range of frequencies below the proton cyclotron 
frequency. These waves scatter particles out of the tails of the initial distributions along 
constant energy surfaces in the wave frame. Overlap of the nonlinear resonance widths allows 
particles to scatter into near isotropic distributions by the end of the simulations. The 
dynamics of $^3$He are explored using test particles. Their temperatures can  increase by a 
factor of nearly 20. Propagation of such waves into regions above and below the flare site can 
lead to heating and transport of $^3$He into the flare acceleration region. The amount of heated 
$^3$He that will be driven into the flare site is proportional to the wave energy. Using values 
from our simulations, we show that the abundance of $^3$He driven into the acceleration region 
should approach that of $^4$He in the corona. Therefore, waves driven by energetic ions produced 
in flares are a strong candidate to drive the enhancements of $^3$He observed in impulsive flares.

\end{abstract}



\section{Introduction} \label{sec:intro}

Solar energetic particles (SEPs) accelerated in impulsive flares provide essential 
clues to the flare energy release process. Due to the lack of {\it in situ} coronal data, 
observations of these events are made through remote sensing of radio, X-ray, gamma-ray, and 
extreme ultraviolet (EUV) emissions. While this has led to progress in understanding the 
acceleration and scattering of energetic electrons in impulsive flares, information on the 
ion dynamics from remote sensing data is limited to flares with significant gamma-ray emission, 
which typically only involve particles with energies in the MeV range and are rarely 
associated with impulsive SEPs \citep{shih_rhessi_2009,lin_energy_2011}. As a result, {\it in 
situ} measurements of the spectra of energetic ions in the solar wind have been essential in 
exploring flare-driven ion acceleration 
\citep{reames_particle_1999,reames_solar_2021,mason_3he-rich_2007,bucik_3he-rich_2020}. 

A characteristic feature of impulsive events is the strong enhancement of the 
$^3$He/$^4$He abundance ratio at high energies, which in some cases is a factor of $10^4$ greater 
than the typical ratios observed in the quiet-time corona and solar wind 
\citep{mason_3he-rich_2007,reames_particle_1999}. This preferential acceleration of $^3$He is 
assumed to be associated with cyclotron resonance interactions due to the unique 
charge-to-mass ratio of this species, an idea first proposed by \cite{fisk_3he-rich_1978}. 
The cyclotron frequency of fully stripped $^3$He (2/3 of the proton cyclotron frequency) is 
distinct from that of other ions in the coronal environment. As a result, waves generated near 
its cyclotron frequency will be preferentially absorbed by $^3$He, while waves around the 
$^4$He cyclotron frequency may be absorbed by other ion species, thus raising the $^3$He/$^4$He 
ratio at high temperatures.

\cite{fisk_3he-rich_1978} proposed a two-stage mechanism whereby electron-ion drifts produced 
during reconnection excite electrostatic ion cyclotron waves that preferentially heat $^3$He 
through its cyclotron resonance. A secondary process with some minimum threshold 
velocity would then accelerate the $^3$He to observed energies. Unfortunately, to produce 
waves of the correct frequency range, \cite{fisk_3he-rich_1978} requires at least 20\% $^4$He 
number density, an abundance much higher than what is typically observed in the corona 
\citep{gabriel_spacelab_1995,moses_global_2020}.

\cite{temerin_production_1992} and \cite{roth_enrichment_1997} propose a different mechanism 
which does not require such a high $^4$He abundance. They suggest reconnection-driven electron 
jets produce oblique electromagnetic cyclotron waves at frequencies just below the proton 
cyclotron frequency. As these waves propagate into regions of stronger magnetic field in the 
lower corona, their frequency remains constant while the local cyclotron frequencies of the 
coronal ion species increase. Since $^3$He has the next highest cyclotron frequency after 
hydrogen, it will resonate with these waves first and therefore preferential acceleration can 
occur. Additional heating of partially-stripped heavy ions could also occur due to 
resonances at the second or higher harmonics. This model has recently been used to explain 
simultaneous $^3$He and ultra-heavy ion enhancements observed by Solar Orbiter 
\citep{mason_heavy-ion_2023}. However, while there is evidence from terrestrial observations that 
cyclotron waves are generated by auroral electrons, the growth rates for this instability are 
small for coronal values of the electron beam energy and $\omega_{pe}/\Omega_{pe}$ 
\citep{temerin_electromagnetic_1984}. Thus, the source of the waves capable of accelerating $^3$He 
remains an open question. 

Both of these proposed mechanisms, along with several others 
\citep[e.g.,][]{zhang_solar_1999,paesold_acceleration_2003} have assumed that 
flare-accelerated electrons are the source of free energy driving the waves responsible for $^3$He 
acceleration. However, simulations show that ions gain more energy than electrons during 
magnetic reconnection \citep{eastwood_energy_2013} and recent observations from Parker Solar 
Probe have found waves in the proton cyclotron frequency band associated with ion beams in the 
solar wind \citep{verniero_parker_2020}. We propose a mechanism for $^3$He acceleration that 
combines the ideas of \cite{fisk_3he-rich_1978} and \cite{temerin_production_1992} but takes 
flare-accelerated ions, rather than electrons, as the wave driver. To explore this hypothesis, we 
perform initial-value simulations to identify and analyze the effects of instabilities 
that form due to high-energy ion populations streaming away from flare sites. We begin with a 
brief review of linear ion instabilities and associtated particle scattering in 
Sec.~\ref{sec:review}.  In Sec.~\ref{sec:param}, we describe the simulation set-up and parameters. 
Results of wave evolution, particle scattering, and $^3$He heating are presented in 
Sec.~\ref{sec:results}. Finally, we end with a discussion of the implications for $^3$He 
acceleration and a summary of conclusions in Sec.~\ref{sec:summary}. 

\section{Background} \label{sec:review}
\subsection{Linear ion instabilities}

Linear ion instabilities have previously been explored in studies of interplanetary shocks 
\citep{gary_electromagnetic_1985,wilson_iii_low_2016,winske_diffuse_1984} and the near-Earth 
solar wind \citep{gary_proton_1976}. From this work, there are five instabilities which could
be relevant for our simulations: the left-hand resonant, right-hand resonant, and non-resonant 
instabilities excited by ion beams streaming parallel to the magnetic field; and the parallel 
and oblique firehose instabilities associated with temperature anisotropies of the form 
$T_{\parallel}/T_{\perp} > 1$, where the subscripts denote direction with respect to the 
magnetic field.

For this discussion, an instability is considered resonant with a particle population if there 
are a sufficient number of particles with parallel velocities that satisfy the resonance 
condition:
\begin{equation}
    \omega - k_{\parallel}v_r \pm n\Omega_{ci} = 0
    \label{eq:resonance}
\end{equation}
where $\omega$ is the wave frequency, $k_{\parallel}$ is the field-aligned component of the 
wavevector, $\Omega_{ci}$ is the cyclotron frequency of the particle species, and the sign of 
the final term depends on the polarization of the wave (positive for right-handed waves and 
negative for left-handed waves). For parallel propagating waves, there is only the n = 1 
resonance. In this case, the resonant velocity $v_r$ corresponds to the frame where the 
Doppler-shifted frequency of the wave matches the cyclotron frequency of the particle. 
Particles with parallel velocities equal to the resonant velocity will experience a constant 
electric field and can coherently exchange energy with the wave. In the oblique case, there 
are an infinite number of resonances corresponding to integer multiples of the cyclotron 
frequency ($n = 0, \pm1, \pm2, ...$). 

\cite{gary_electromagnetic_1984}, \cite{gary_electromagnetic_1991}, and 
\cite{gary_theory_1993} provide an extensive review of the instabilities caused by 
counter-streaming ion populations. Assuming simplified velocity distributions, where both beam 
and core ions are represented by drifting Maxwellians, there are three modes which can be 
excited based on the beam temperature, density, and drift speed.

The right-hand circularly polarized mode, emerging from the magnetosonic branch of the 
dispersion relation, is the dominant mode for cool, tenuous beams with low-to-moderate drift 
speeds. This mode propagates parallel to the magnetic field with a phase velocity 
$v_p \approx v_A$ and frequencies $\omega \lesssim \Omega_{ci}$. As the temperature of the 
beam is increased, a left-hand circularly polarized mode from the Alfv\'en/ion cyclotron 
branch is also driven unstable. Like the right-hand mode, it propagates parallel to the 
magnetic field with $v_p \approx v_A$ and $\omega \lesssim \Omega_{ci}$. Both modes are 
resonant with the beam ions and thus we refer to them as the right-hand resonant and left-hand 
resonant instabilities. For cool, fast beams with considerable densities, the dominant mode is 
purely-growing ($\omega \approx 0$) and non-resonant for both the core and beam ion
populations. We refer to this as the non-resonant instability. 

In addition to ion beam instabilities, we also consider instabilities induced by the 
temperature anisotropy $T_{\parallel}/T_{\perp} > 1$. In the MHD limit, this produces the 
purely-growing fire hose instability with the threshold 
$\beta_{\parallel} - \beta_{\perp} = 2$. On kinetic scales, assuming a bi-Maxwellian 
distribution, this mode has frequencies $\omega \lesssim \Omega_{ci}$ and it propagates 
parallel to the magnetic field with a phase velocity $v_p \gtrsim v_A$ and a right-handed 
polarization \citep{ kennel_limit_1966,gary_theory_1993}. As shown in \cite{gary_proton_1976}, 
the threshold for instability is also slightly reduced in this regime.  While the classic fire 
hose instability emerges from the magnetosonic branch of the dispersion relation, 
\cite{hellinger_new_2000} found that the Alfv\'en branch can also be driven unstable by the 
same anisotropy, producing a purely-growing and strongly oblique instability. 

\subsection{Resonance widths and quasi-linear particle diffusion}
For parallel propagating waves, the electric field goes to zero in the reference frame moving with the wave. 
Therefore, wave-particle interactions in this frame conserve energy. For the non-relativistic case we 
consider in this paper, this constrains resonant particle motion in $v_{\parallel}$-$v_{\perp}$ phase 
space to concentric circles centered at the phase speed of the wave. How far particles are able to 
scatter along these surfaces depends on the nonlinear resonance widths, which were derived in 
\cite{karimabadi_particle_1990} using formal Hamiltonian theory. For low amplitude waves, marginal 
stability can be reached when the gradients along the resonant portions of the distribution function 
have been reduced to zero \citep{kennel_velocity_1966}. However, in the case of oblique waves or 
multiple modes of varying wavenumbers and phase speeds, large amplitudes can result in significant 
resonance overlap, allowing for open phase space orbits corresponding to arbitrarily high energies 
\citep{karimabadi_physics_1992}. In this case, particles will be continuously heated until the waves are 
damped or until the wave amplitude has decreased to a point where the resonances no longer overlap.

\section{Simulation Parameters} \label{sec:param}

We perform fully kinetic, 2.5D simulations using the particle-in-cell (PIC) code {\tt p3d} 
\citep{zeiler_three-dimensional_2002}, in which we follow the self-consistent dynamics of three particle 
species: electrons, protons, and alpha particles. The number density of the alphas is $5\%$ 
that of the protons and additional electrons are added in order to balance the charge density. Due to 
the low abundance of $^3$He in the corona, we assume that this population 
has a negligible effect on the dynamics of the system and thus include it as a population of 
test particles.

We use normalized units in our simulations. All lengths are normalized to the 
proton inertial length $d_p$, all times to the inverse of the proton cyclotron frequency 
$\Omega_{cp}^{-1}$, all speeds to the proton Alfv\'en speed $v_A$, and all temperatures to 
$m_pv_A^2$. In order to reduce the computational cost, we take the standard steps of using 
non-physical values for the speed of light, $c = 20v_A$, and the electron-to-proton mass ratio, 
$m_e/m_p = 1/25$.

We enforce periodic boundary conditions in both the $x$- and 
$y$-directions and assume no variation in the $z$-direction. Velocities 
and fields have been calculated with three spatial components, but in a two-dimensional spatial
domain with $L_x = 102.4 d_p$ and $L_y = 25.6 d_p$. We use a grid spacing of 
$\Delta x = 0.0125d_p$, giving 8192 x 2048 grid points in total, and use 50 particles 
per cell for the protons.

All simulations begin with a uniform magnetic field $\mathbf{B} = B_0\mathbf{\hat{x}}$ 
and no electric field. Initial distribution functions for the protons and alphas consist 
of two populations of equal density: a cold ($v_{th} < v_A$) Maxwellian background 
with $T_c = 0.1 m_pv_A^2$ and a hot ($v_{th} > v_A$) streaming population represented 
by an anisotropic, one-sided kappa function:
\begin{equation}
    f_\kappa = \frac{\Gamma(\kappa + 1)}{\pi^{3/2}\theta^2_\perp\theta_x\kappa^{3/2}
    \Gamma(\kappa - 1/2)}\left[1 + \frac{v^2_x}{\kappa\theta^2_x} + 
    \frac{v^2_\perp}{\kappa\theta^2_\perp}\right]^{-(\kappa + 1)}\Theta(v_x)
    \label{eq:distfunc}
\end{equation}
where $\Gamma$ is the gamma function, $\Theta$ is the Heaviside step function, and
$\theta_i^2 = (2T_i/m_i)*[(\kappa - 3/2)/\kappa]$. For all 
simulations, we use $\kappa = 2.5$ and an initial temperature anisotropy of 
$T_{\kappa,x}/T_{\kappa,\perp} = 10$.

This combined distribution models a population of accelerated ions that exhibits the non-thermal 
power-law tails often observed in flares \citep{reames_energy_1997,mason_3he-rich_2007} 
and is streaming out from a reconnection site into a cold background plasma. The equal density of 
the energetic population and cold ion background reflects the high efficiency of particle 
acceleration that is expected fromreconnection \citep{arnold_electron_2021}, while the large value 
of $T_{\kappa,x}/T_{\kappa,\perp}$ is a consequence of Fermi reflection, which dominantly increases 
energy parallel to the ambient magnetic field \citep{drake_electron_2006,dahlin_mechanisms_2014}. 

We perform three runs with differing values of the initial energy of the streaming population: Run 
1 has an initial $T_{\kappa,x} = 10 m_pv_A^2$, while Run 2 is less energetic with an initial 
$T_{\kappa,x} = 5 m_pv_A^2$ and Run 3 is more energetic with an initial 
$T_{\kappa,x} = 20 m_pv_A^2$. For all runs, the initial distribution function of the electrons 
is a Maxwellian with $T_e = 2 m_pv_A^2$. A small electron drift is also added to balance the 
current. The fully ionized $^3$He test particles are initialized as a Maxwellian with the same
temperature as the cold ion background, $T_c = 0.1 m_pv_A^2$. 

\section{Simulation Results}\label{sec:results}
\subsection{Waves induced by the initial distribution function}
\begin{figure}[ht]
\plotone{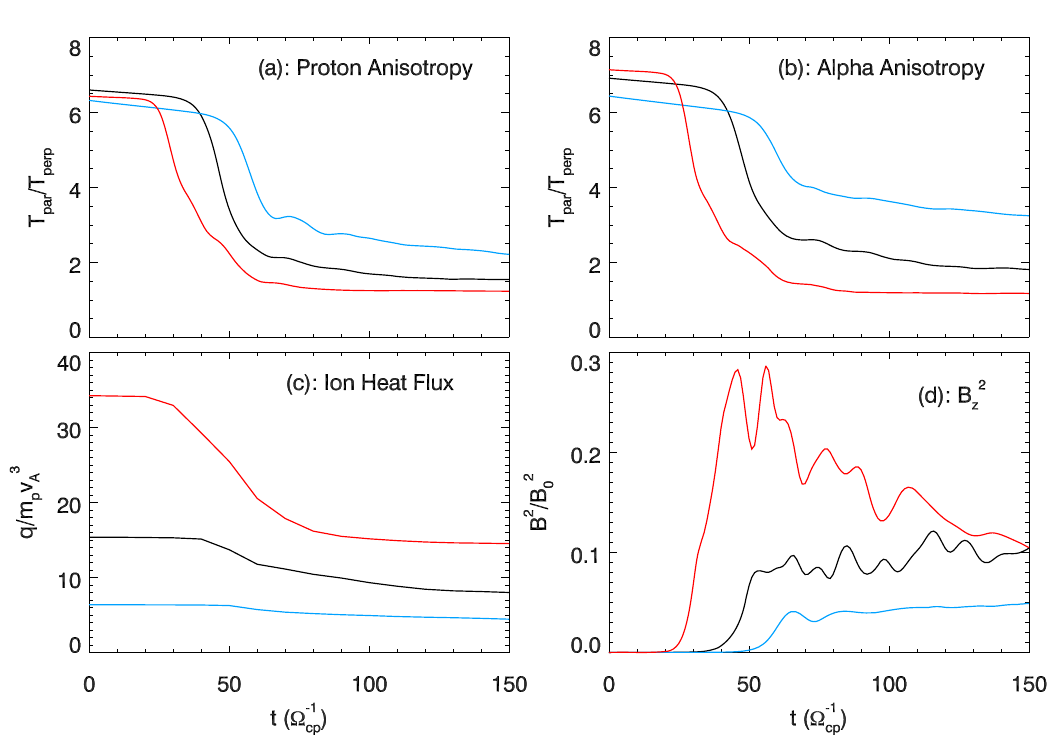}
\caption{(a) Proton $T_{\parallel}/T_{\perp}$, (b) alpha $T_{\parallel}/T_{\perp}$, (c) combined 
proton and alpha heat flux, and (d) perpendicular magnetic energy versus time for each 
case. Run 1 ($T_{\kappa,x} = 10 m_pv_A^2$) is plotted in black, Run 2 ($T_{\kappa,x} = 5 m_pv_A^2$) is 
plotted in blue, and Run 3 ($T_{\kappa,x} = 20 m_pv_A^2$) is plotted in red. The ion temperatures and 
perpendicular magnetic field data are output directly by the code every $1 \Omega_{cp}^{-1}$. The 
quantities shown here are the averages over the whole domain at each time. The heat fluxes are 
calculated using the 3D distribution functions and therefore data is only available every 10 
$\Omega_{cp}^{-1}$. 
\label{fig:anisotropy}}
\end{figure}

Simulations are run to a total time $t = 150 \Omega_{cp}^{-1}$. Evolution of 
the proton and alpha temperature anisotropies, the combined proton and alpha parallel heat 
flux, and the out-of-plane magnetic field energy are presented in Figure \ref{fig:anisotropy}. 
All three sets of initial conditions produce instabilities driven by both the temperature 
anisotropies and the parallel ion heat flux, with growth rates and peak perpendicular magnetic 
energies that increase as the initial energy of the streaming population is increased. Due to the 
complexity of the initial distribution functions, we have not analytically solved the linear 
dispersion relation. Therefore, we first analyze the characteristics of the waves measured in the 
simulations and then compare our results to the instabilities discussed in Sec.~\ref{sec:review}.

For Run 1, instability onset occurs around $t = 35 \Omega_{cp}^{-1}$. The wave amplitude peaks and 
stabilizes at $t = 50 \Omega_{cp}^{-1}$ with the perpendicular magnetic energy equal to around 10\% 
that of the initial magnetic energy. By $t = 60 \Omega_{cp}^{-1}$, the anisotropies begin plateauing 
to their final values, while the heat flux continues to steadily decrease until $t = 130 
\Omega_{cp}^{-1}$.

Decreasing the initial beam energy in Run 2 delays the time of instability onset, in this case 
occurring at $t = 50 \Omega_{cp}^{-1}$. Wave amplitude peaks and again stabilizes at $t = 65 
\Omega_{cp}^{-1}$ with the perpendicular magnetic field energy equal to half that seen in Run 1. 
The final values of both anisotropies and the heat flux are higher than those of Run 1, as expected 
from the lower energy fluctuations. 

Increasing the initial beam energy in Run 3 leads to an instability which develops earlier in 
time, at $t = 20 \Omega_{cp}^{-1}$, and is significantly more energetic. Wave energy peaks at 
$t = 45 \Omega_{cp}^{-1}$ with a perpendicular magnetic energy equal to nearly 30\% of the initial 
magnetic energy. Unlike the two less energetic cases, some of this wave energy is then dissipated 
throughout the simulation, ending with a final value equal to that of Run 1. There is a large 
reduction in both the heat flux and the anisotropies, with both the protons and alphas ending 
the run nearly isotropic in temperature.

\begin{figure}[h!]
\plotone{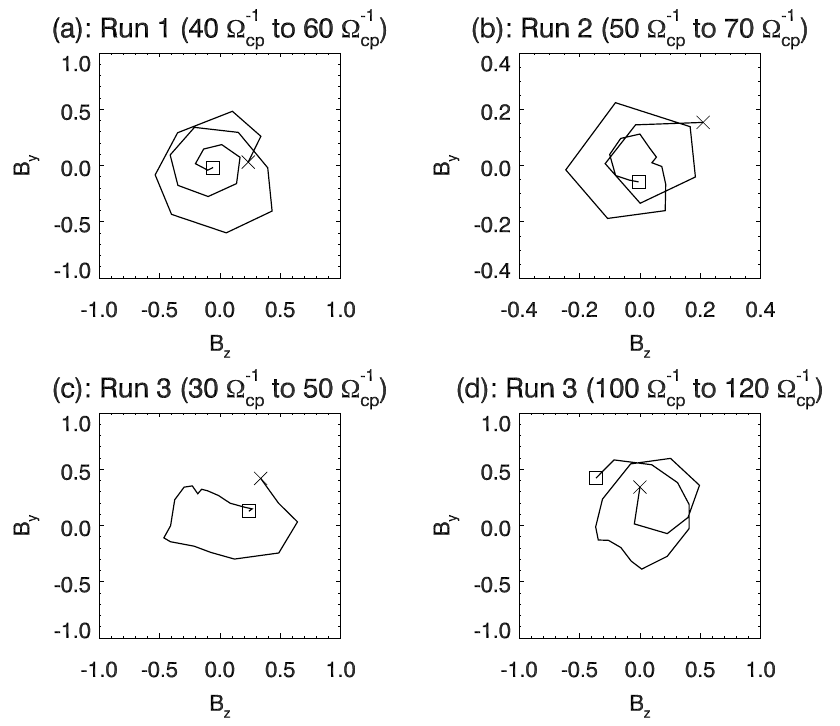}
\caption{Hodograms ($B_z$ vs. $B_y$) at $x = 25 d_p$, $y = 12.5 d_p$, every $1 \Omega_{cp}^{-1}$ 
for each case. Initial times are indicated by the box, while final times are indicated by the X. The 
first three plots contain data around the peak of each instability: (a) 40 to 60 $\Omega_{cp}^{-1}$ 
for Run 1, (b) 50 to 70 $\Omega_{cp}^{-1}$ for Run 2, and (c) 30 to 50 $\Omega_{cp}^{-1}$ for Run 3. 
An additional plot with data near the end of Run 3 (100 to 120 $\Omega_{cp}^{-1}$) is included to show 
that the instability changes polarization at late time in the simulation. 
\label{fig:polarization}}
\end{figure}

\begin{figure}[h!]
\plotone{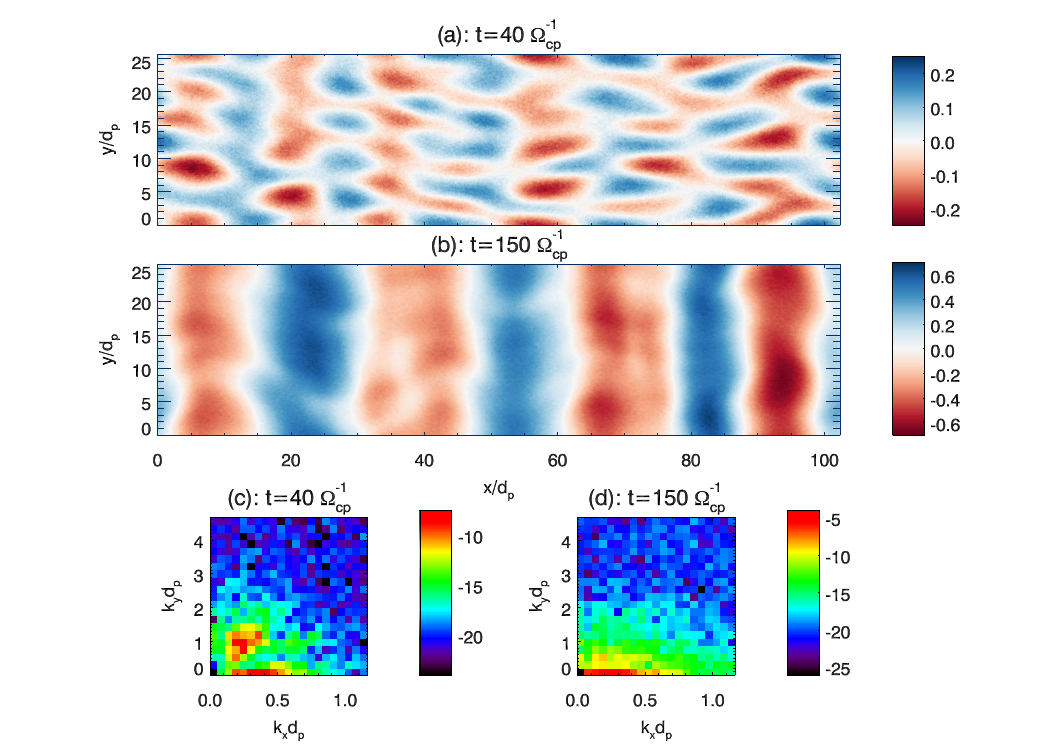}
\caption{(a-b) Out-of-plane magnetic field data and (c-d) the natural logarithm of the FFT power 
spectra for Run 1 at early and late time. The spectra peak at $k_yd_p = 0$ for both times, with 
$k_xd_p = 0.37$ at $t = 40 \Omega_{cp}^{-1}$ and $k_xd_p = 0.18$ at $t = 150 \Omega_{cp}^{-1}$. 
\label{fig:fft}}
\end{figure}

\begin{figure}[h!]
\plotone{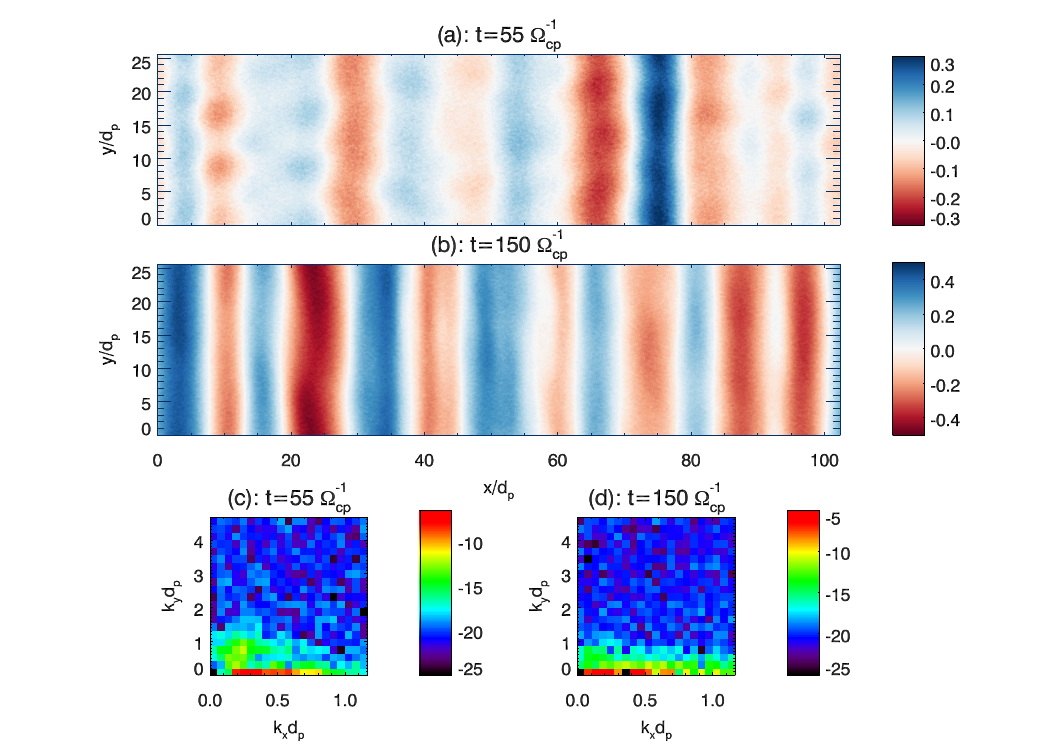}
\caption{(a-b) Out-of-plane magnetic field data and (c-d) the natural logarithm of the spatial FFT 
power spectra for Run 2 at early and late time. The spectra peak at $k_xd_p = 0.37$ and 
$k_yd_p = 0$ for both times.
\label{fig:fftcold}}
\end{figure}

\begin{figure}[h!]
\plotone{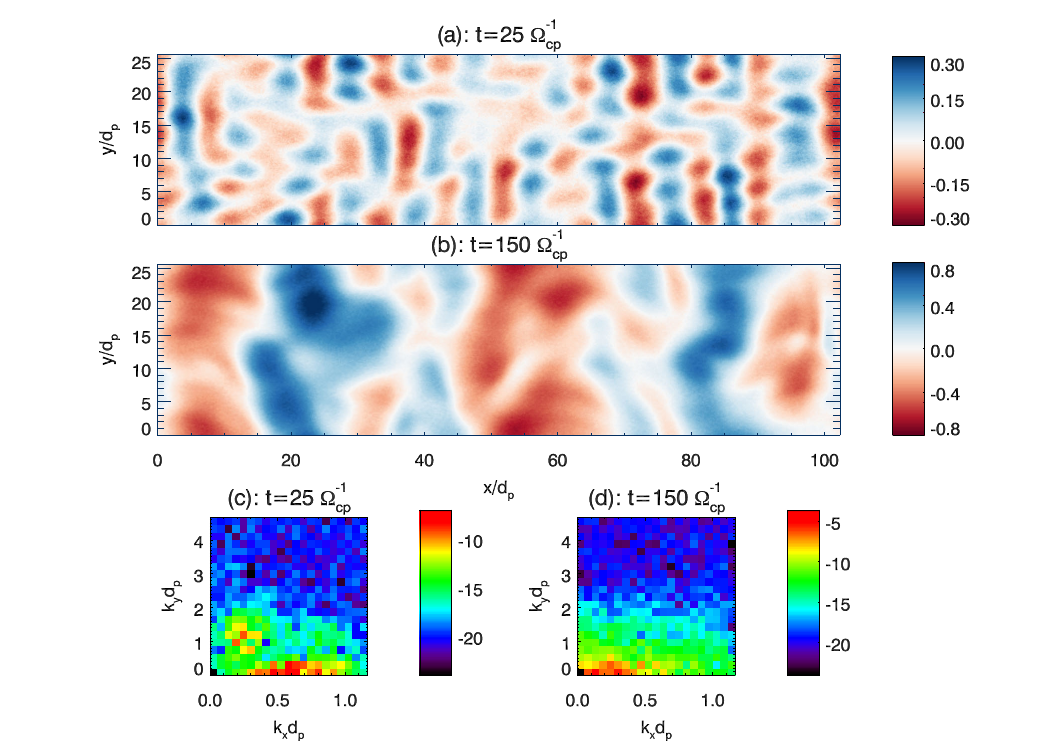}
\caption{(a-b) Out-of-plane magnetic field data and (c-d) the natural logarithm of the spatial FFT 
power spectra for Run 3 at early and late time. The spectra peak at $k_yd_p = 0$ for both times, 
with $k_xd_p = 0.67$ at $t = 25 \Omega_{cp}^{-1}$ and $k_xd_p = 0.12$ at $t = 150 \Omega_{cp}^{-1}$.
\label{fig:ffthot}}
\end{figure}

\begin{figure}[h!]
\plotone{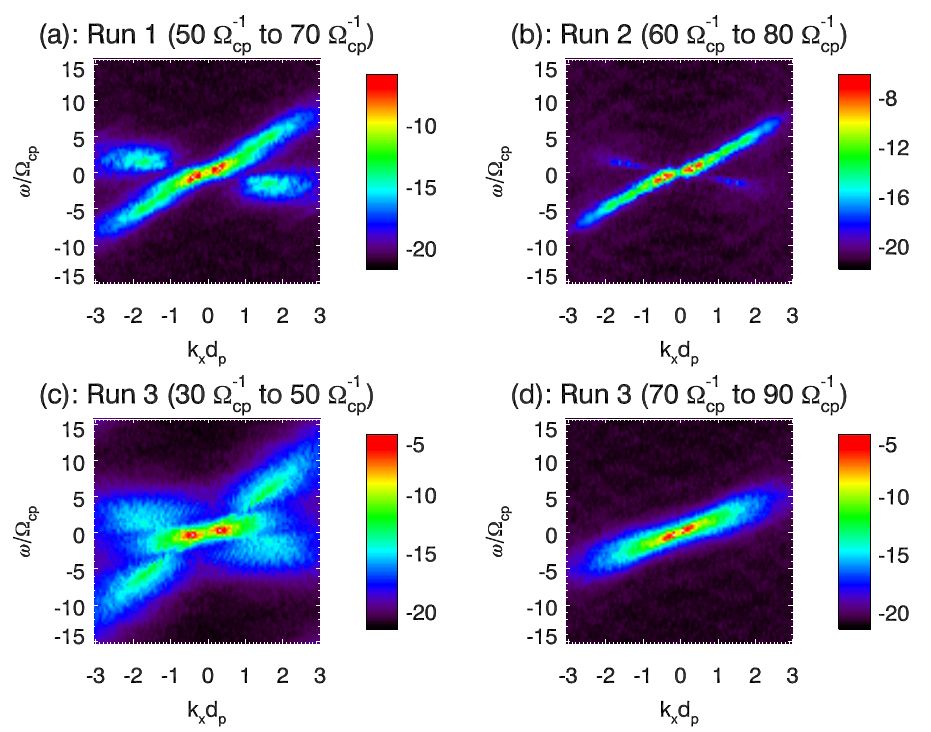}
\caption{The natural logarithm of the power spectra for FFTs along the parallel direction in space and 
over 20 $\Omega_{cp}^{-1}$ in time for (a) Run 1 at t = 50 to 70 $\Omega_{cp}^{-1}$, (b) Run 2 at t = 
60 to 80 $\Omega_{cp}^{-1}$, (c) Run 3 at t = 30 to 50 $\Omega_{cp}^{-1}$, and (d) Run 3 at t = 70 to 
90 $\Omega_{cp}^{-1}$. Each plot is the average of 1,024 FFTs taken along each horizontal cut in the 
lower half of the domain. Peaks occur at (a) $k_xd_p = 0.37$, $\omega/\Omega_{cp} = 0.63$, (b) 
$k_xd_p = 0.37$, $\omega/\Omega_{cp} = 0.63$, (c) $k_xd_p = 0.37$, $\omega/\Omega_{cp} = 0.31$, and 
(d) $k_xd_p = 0.12$, $\omega/\Omega_{cp} = 0.31$.
\label{fig:omegakpar}}
\end{figure}

With an understanding of the temporal evolution of the instabilities and drivers, we next look
at the characteristics of the waves they produce. We calculate the wave polarizations for each 
run by plotting $B_z$ versus $B_y$ at a single point in the domain for 20 points in time spaced every 
$1 \Omega_{cp}^{-1}$. These results are presented in Figure \ref{fig:polarization}. We then find 
the spectrum of unstable parallel ($k_{x}$) and perpendicular ($k_{y}$) wavenumbers by taking fast 
Fourier transforms (FFTs) of the out-of-plane magnetic field fluctuations over the whole domain. 
These results are presented in Figures \ref{fig:fft}-\ref{fig:ffthot}. The phase 
velocities are calculated directly by comparing parallel and perpendicular cuts of the out-of-plane 
magnetic field data at multiple consecutive times. Finally, additional FFTs are taken spatially over 
the parallel direction and temporally over $20 \Omega_{cp}^{-1}$ to produce the $k_x$ versus $\omega$ 
spectra presented in Figure \ref{fig:omegakpar}. In order to reduce noise, each spectrum is
averaged over the lower half of the spatial domain and calculated using the perpendicular 
magnetic field data multiplied by the function $sin^2(\pi (t-t_0)/T)$, where $t_0$ is the initial 
time and $T$ is the total time period, ensuring periodicity in time.

The initial conditions of Run 1 excite right-hand polarized waves (Fig.~\ref{fig:polarization}a). 
From the 2D plot of $B_z$ in Fig.~\ref{fig:fft}a and the FFT in Fig.~\ref{fig:fft}c, we see two 
modes at early time: a short wavelength, perpendicular mode ($k_yd_p \approx 1, k_x \ll k_y$) and a 
longer wavelength parallel mode ($k_xd_p < 1, k_y \ll k_x$), both with a phase velocity 
$v_p \approx 2v_A$ in the parallel direction. As time progresses, the perpendicular mode dissipates 
and by the end of the simulation we are left with predominantly parallel waves with increased 
wavelengths (see Fig.~\ref{fig:fft}d).

In the $k_x$ versus $\omega$ spectrum (Fig.~\ref{fig:omegakpar}a), the majority of the wave power is 
in frequencies at and below the proton cyclotron frequency with corresponding parallel wavenumbers 
$k_xd_p \leq 0.5$, consistent with forward propagating waves moving at $2v_A$. At higher wavenumbers 
and frequencies, the power from this mode decreases before dropping off around $k_xd_p = 1$ and 
$\omega = 2 \Omega_{cp}$. Two additional weak modes exist above $k_xd_p = 1$, with both positive and 
negative frequencies greater than $\Omega_{cp}$, corresponding to forward and backward propagating 
modes.

In the less energetic Run 2, the perpendicular mode is significantly weaker and dissipates 
much quicker than in Run 1, thus the instability has very little perpendicular structure 
throughout the simulation (Fig.~\ref{fig:fftcold}a and Fig.~\ref{fig:fftcold}c). In addition, this 
case has higher wavenumbers at late time (Fig.~\ref{fig:fftcold}b and Fig.~\ref{fig:fftcold}d) and 
the backward propagating mode that appears in the $k_x$ versus $\omega$ spectrum is weaker 
(Fig.~\ref{fig:omegakpar}b). Apart from this, the characteristics remain similar to those seen in Run 
1: the waves are right-hand polarized (Fig.~\ref{fig:polarization}b), with a parallel phase speed 
$v_p \approx 2v_A$, frequencies up to the proton cyclotron frequency, and small wavenumbers 
($k_xd_p < 1$).

Raising the initial energy in Run 3 once again produces a short wavelength, perpendicular mode 
with wavenumbers similar to those seen in Run 1 (Fig.~\ref{fig:ffthot}a and Fig.~\ref{fig:ffthot}c). 
However, the parallel mode at early time is significantly different, with larger wavenumbers, a 
left-handed polarization (Fig.~\ref{fig:polarization}c), and a much slower phase speed 
($v_p < v_A$), corresponding to lower frequencies: $\omega < 0.5 \Omega_{cp}$ 
(Fig.~\ref{fig:omegakpar}c). Apart from this, the $k_x$ versus $\omega$ spectrum at early time is 
similar to those of the previous two cases, albeit more diffuse due to the changing wavenumbers 
during this period.

Later in the run, the perpendicular mode weakens and the parallel 
mode begins to resemble those of the other two cases: the polarization becomes right-handed 
(Fig.~\ref{fig:polarization}d) and the wavenumbers decrease (Fig. ~\ref{fig:ffthot}d) while the 
frequency remains roughly constant (Fig.~\ref{fig:omegakpar}d), leading to an increase in the phase 
velocity. The higher frequency modes also disappear at this time, leaving only the primary mode with 
peak frequencies below the proton cyclotron frequency and $kd_p < 0.5$.

While our initial distribution functions are far from the simple Maxwellians considered in 
the literature (see discussion in Sec.~\ref{sec:review}), it may still be useful to compare our 
results to the linear instabilities previously discussed. To do this, we shift to the rest frame of 
the full proton and alpha distributions. This corresponds to a frame moving parallel to the 
magnetic field at $\approx 1v_A$ for Run 1, $\approx 0.75v_A$ for Run 2, and $\approx 1.5v_A$ for 
Run 3. In this frame, the parallel mode in Run 1, Run 2, and at late-time in Run 3 is a forward 
propagating right-handed wave with a phase speed $v_p \approx v_A$ and frequencies around 
$0.3\Omega_{cp}$. It is most likely a form of the right-hand resonant instability or possibly a 
combination of the right-hand resonant and parallel fire hose instabilities. At early-time in Run 
3, this mode has similar characteristics, but is backward propagating, unlike any of the 
instabilities discussed in Sec.~\ref{sec:review}. 

The oblique mode appears to have an energy threshold for instability, as it only appears in the 
higher energy cases of Runs 1 and 3, though it is unclear how it relates to the linear instabilities 
previously discussed. Likewise, we have not identified the high frequency modes that appear in the 
$k_x$ versus $\omega$ spectra; however, given that they are considerably weaker than the dominant 
mode and do not appear in the $k_x$ versus $k_y$ spectra, it is not likely that they play a 
significant role in the simulation dynamics.

\subsection{Proton and alpha wave-particle interactions}

\begin{figure}[h!]
\plotone{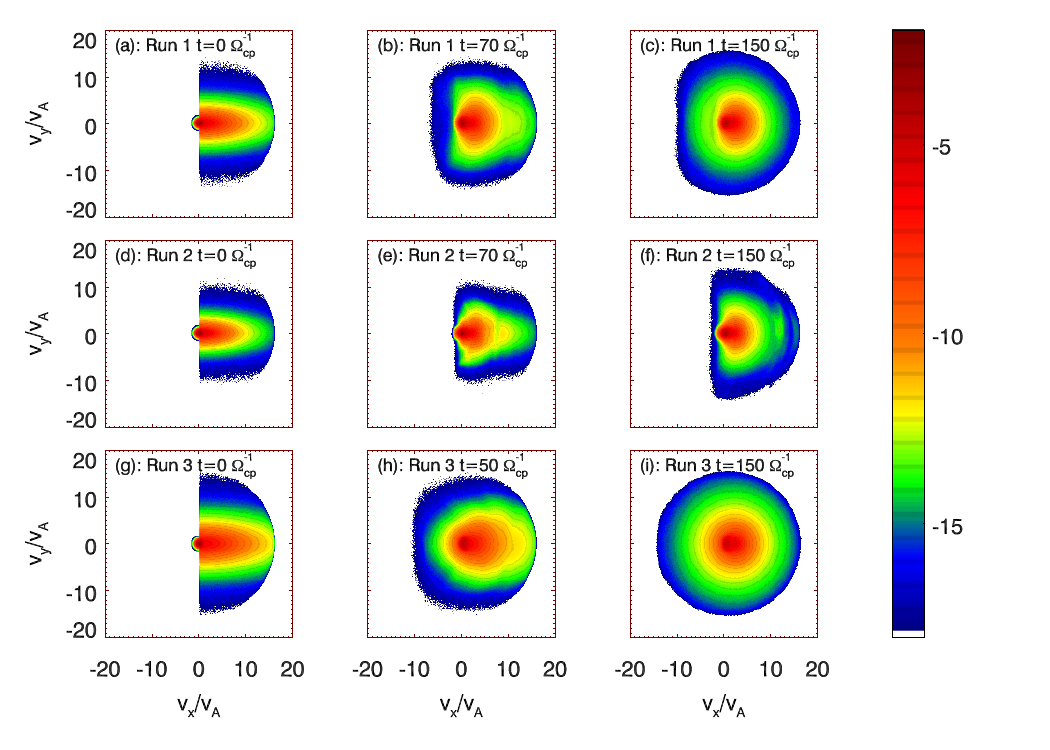}
\caption{Velocity distribution functions for the protons at the beginning, middle, and end of Run 1 
(a-c), Run 2 (d-f), and Run 3 (g-i). Each distribution is plotted on the same logarithmic scale and 
is normalized to 1. 
\label{fig:vdist}}
\end{figure}

To study the wave-particle interactions driving these instabilities, we generate velocity 
distribution functions for the protons and the alphas every 10 $\Omega_{cp}^{-1}$. Because $B_x$ remains 
the dominant magnetic field component for the duration of each simulation, the velocities
$v_y$ and $v_z$ remain the dominant perpendicular components with $v_x$ the parallel component. For 
simplicity we display the $v_x-v_y$ velocity distribution functions to illustrate the relaxation of the 
initial anisotropic distribution functions as the instabilities develop. Proton 
distributions from the beginning, middle, and end of each run are presented in Figure 
\ref{fig:vdist}. The dynamics of the alpha particles are similar to those of the protons in all 
cases and therefore are not shown here.

In Run 2, where there are no oblique modes, wave-particle interactions occur only at the $n = +1$ 
resonance or $v_\parallel=(\omega +\Omega_{cp})/k_\parallel = 2v_A+\Omega_{cp}/k_\parallel$. 
Particles from the high-energy tail of the distribution scatter along circular orbits 
centered at $v_x = 2v_A$, corresponding to constant energy surfaces in the frame of the wave. The 
lower energy particles scatter first and contribute to the growth of the instability. As time 
progresses and the parallel wavenumbers decrease, the higher energy particles become resonant. 
However, due to the fewer number of particles at high energies, the total energy transferred to the 
waves by these particles is relatively low. Without multiple resonances, particle scattering in this 
case is limited and there remains a clear drop-off in the distributions around $v_x = 0$. This is 
consistent with the quasilinear theory for parallel waves, wherein particles are unable to scatter 
past 90 degrees in pitch angle. 

\begin{figure}[h!]
\plotone{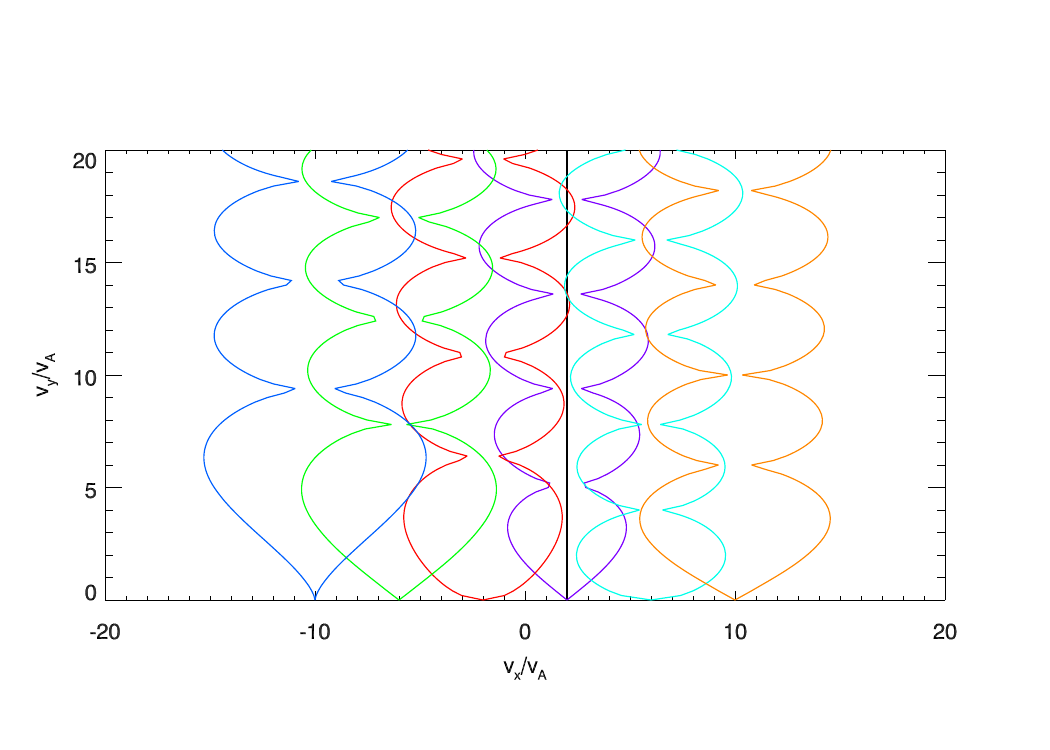}
\caption{Proton resonance widths at t = 70 $\Omega_{cp}^{-1}$ for Run 1, using 
$k_{\parallel} = 0.25d_p$ and $k_{\perp} = 0.75d_p$. The phase speed of the wave is plotted 
in black, while the $n = -3$ (blue), $n = -2$ (green), $n = -1$ (red), $n = 0$ (purple), $n = 1$ 
(cyan), and $n = 2$ (orange) resonances are plotted as the area enclosed by each curve. 
Wavenumbers were selected by taking the strongest oblique mode from the FFT power spectrum 
and the fluctuation amplitude used is the root-mean-square average over the box.
\label{fig:pres}}
\end{figure}

In Runs 1 and 3, where there are finite amplitude oblique modes, a variety of resonances ($n = 0$, 
$\pm1$, $\pm2$, and $\pm3$) overlap and efficiently scatter particles from positive to 
negative parallel velocities. An example of this is given in Fig.~\ref{fig:pres}, where we have 
calculated the resonance widths for Run 1 using Equations 5a-5g from \cite{karimabadi_physics_1992}, 
assuming circularly polarized waves with positive helicity ($A_1 = A_2 = \delta B/k_{\parallel}$) 
and no electrostatic component ($\Phi_0 = 0$). As in the parallel case, particles in Run 1 scatter 
along circles centered at $v_x = 2v_A$. In Run 3, the dynamics are much more chaotic due to the 
changing phase speed of the waves. In both cases, it is once again the lower energy particles that 
scatter first and contribute the most energy towards wave growth. Along with the broad range of 
wavenumbers in both cases, resonance overlap allows the majority of velocity space to diffuse in 
pitch angle and particles are able to scatter into near isotropic distributions by the end of the 
simulations. This is most apparent in the final distributions for Run 3.

\subsection{$^3$He resonance heating}
\begin{figure}[h!]
\plotone{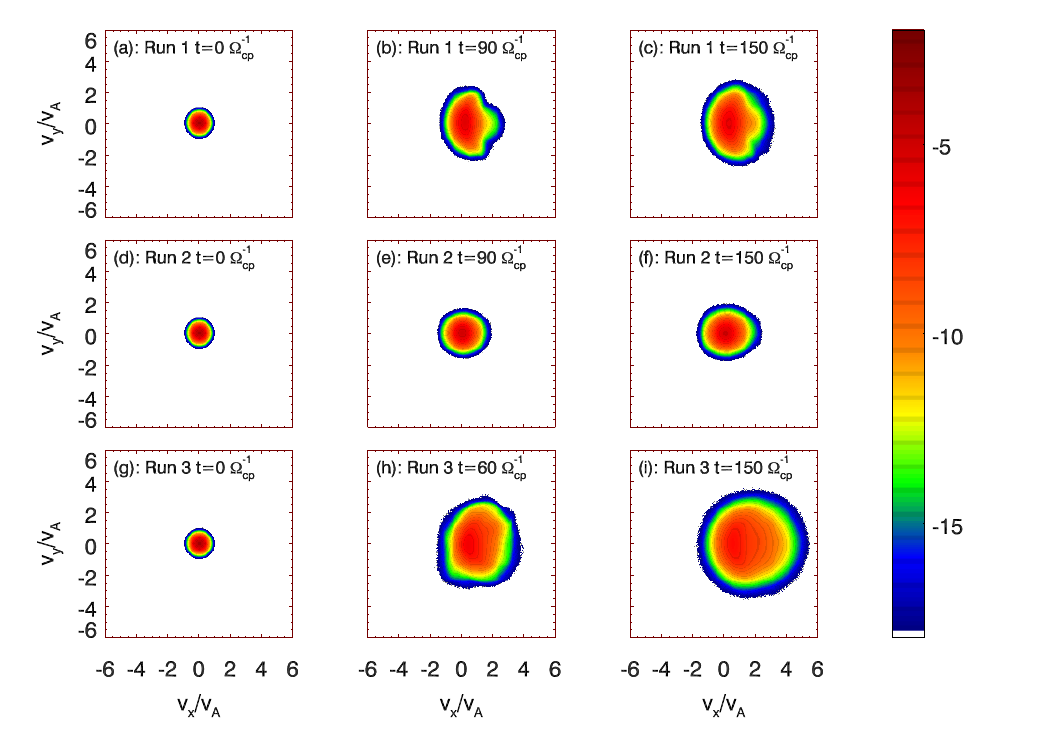}
\caption{Velocity distribution functions of the $^3$He test particles at the beginning, middle, and 
end of (a-c) Run 1, (d-f) Run 2, and (g-i) Run 3. Each distribution is plotted on the same 
logarithmic scale and is normalized to 1.
\label{fig:h3vdist}}
\end{figure}

\begin{figure}[h!]
\plotone{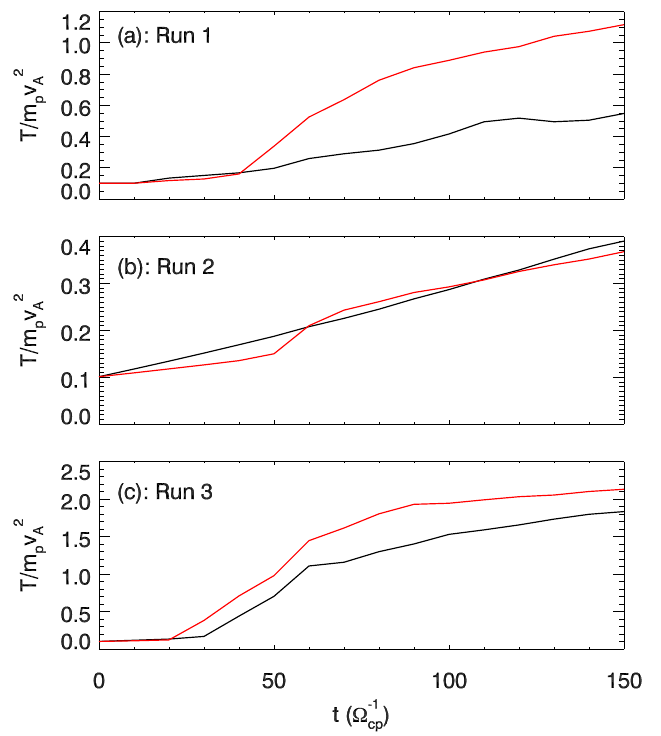}
\caption{Parallel (black) and perpendicular (red) $^3$He temperatures versus time for each run, 
with data plotted every 10 $\Omega_{cp}^{-1}$. Note that the vertical scale changes between each plot.
\label{fig:h3temps}}
\end{figure}

The $^3$He dynamics have been calculated in all three cases; however, as previously stated in 
Sec.~\ref{sec:param}, we neglect the feedback of these particles on the simulation because of their 
low density in the corona. In the same manner as the protons and alphas, we generate velocity 
distribution functions for the $^3$He ions every $10 \Omega_{cp}^{-1}$. Three times from each 
case are presented in Figure \ref{fig:h3vdist}. We also calculate the parallel and perpendicular 
temperatures of the $^3$He directly from the full, three-dimensional distribution functions. The 
temperatures versus time for all three runs are presented in Figure \ref{fig:h3temps}.

In all cases, the resonant velocity at $v_\parallel = (\omega + \Omega_{cp})/k_\parallel$ is much 
greater than the initial thermal velocity of the $^3$He and thus this resonance does not contribute 
to the initial heating of these particles. For Run 2, where this is the only active resonance, this 
means that the $^3$He heating throughout the run is minimal in both the parallel and perpendicular 
directions. 

\begin{figure}[h!]
\plotone{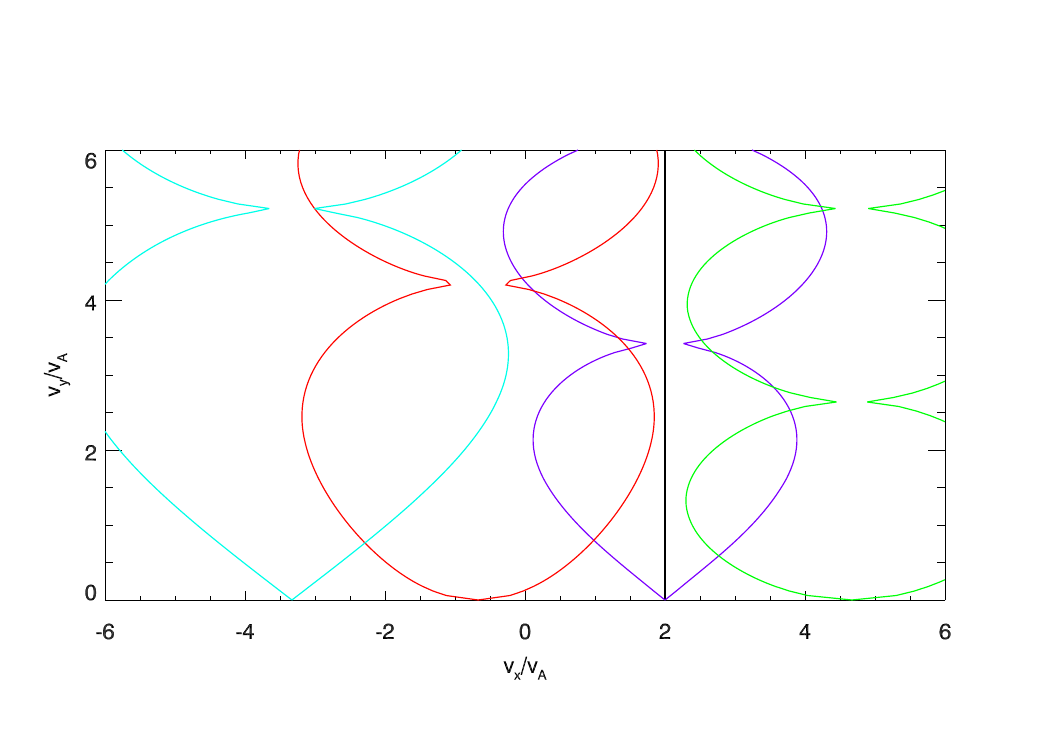}
\caption{$^3$He resonance widths at t = 50 $\Omega_{cp}^{-1}$ for Run 1, using 
$k_{\parallel} = 0.25d_p$ and $k_{\perp} = 0.75d_p$. The phase speed of the wave is plotted 
in black, while the n = -2 (cyan), n = -1 (red), n = 0 (purple), and n = 1 (green) 
resonances are plotted as the area enclosed by each curve. Wavenumbers were selected by 
taking the strongest oblique mode from the FFT power spectrum and the fluctuation amplitude 
used is the root-mean-square average over the box.
\label{fig:h3res}}
\end{figure}

For Run 1, where there are oblique waves at early time, $^3$He is initially heated in the 
perpendicular direction by the $n = -1$ resonance and in the parallel direction by the $n = 0$ 
resonance. This is most clearly demonstrated in Fig.~\ref{fig:h3vdist}b, where positive velocity 
particles are pulled towards the phase speed of the wave, $v_p = 2v_A$, while negative velocity 
particles are pulled upwards and into circular orbits about $v_p$. As time progresses, particles that 
have been heated in the perpendicular direction also gain parallel energy due to resonance 
broadening and overlap of the $n = -1$, 0, and +1 resonances (Figure \ref{fig:h3res}), though the 
majority of the energy gained remains in the perpendicular direction.

The heating and scattering of $^3$He is most effective in Run 3 due to the changing phase speed and 
polarization of the waves. At early time, the waves are left-hand polarized and strong perpendicular 
heating initially occurs through the primary ($n = +1$) resonance at 
$v_\parallel = (\omega - \Omega_{cp})/k_\parallel$. As the phase speed of the wave changes, the 
particle orbits no longer follow the simple concentric circles discussed in Sec.~\ref{sec:review} and 
instead become chaotic, leading to more isotropic heating as reflected in Fig.~\ref{fig:h3vdist}h. 
Once the polarization of the waves switches and the primary resonance moves to 
$v_\parallel = (\omega + \Omega_{cp})/k_\parallel$, additional scattering and parallel heating also 
occurs, allowing the final distribution to more closely resemble a Maxwellian than in the previous 
two cases. 

\section{Discussion and Conclusions}\label{sec:summary}

We have shown, using fully kinetic plasma simulations, that energetic protons and alpha particles 
streaming into a region of colder plasma can generate ion-scale instabilities ($kd_p \lesssim 1$). 
For sufficiently large energies, both parallel and oblique modes are excited, leading to resonances 
at both positive and negative velocities. Resonance overlap occurs as the fluctuation amplitudes 
increase, allowing the energetic particles to scatter out of the tails of the distributions, past 90 
degrees in pitch angle and into negative velocities. For the most energetic initial conditions, this 
scattering is very efficient and results in nearly isotropic distributions by the end of the 
simulation. The scattering time is around 100 $\Omega_{cp}^{-1}$, corresponding to $\approx 10^{-2}$ 
seconds in the corona. For the highest energy ions, this gives a mean free path of $\approx 1$ Mm, 
within the typical size (10's Mm) of the flare energy release region. Therefore, we expect that ions 
with large parallel energies due to reconnection will generate finite amplitude waves that act to 
scatter parallel energy into perpendicular energy and therefore inhibit energetic ion transport out 
of the flare energy release region. 

These waves are also effective at heating trace amounts of $^3$He, increasing both the parallel and 
perpendicular temperatures by a factor of nearly 20 in the most energetic of our simulations. A key 
question, however, is whether these waves are capable of boosting the abundance of $^3$He in impulsive 
flares. In the model by \cite{temerin_production_1992}, left-handed waves driven by electron beams at 
or slightly below the proton cyclotron frequency propagate into regions of higher magnetic field. As 
these waves propagate, their frequency remains constant, while the ion cyclotron frequencies increase 
proportional to the increasing field. The $^3$He resonance condition for these waves is given by
$v_\parallel = (\omega - \Omega_{ci})/k_\parallel$, where $\Omega_{ci} = (2/3)\Omega_{cp}$ is the 
$^3$He cyclotron frequency. Because the ambient $^3$He outside of the flare energy release region is 
cold, the resonance interaction requires $\omega\sim\Omega_{ci}$. Since $^3$He has the highest 
cyclotron frequency of all of the non-protons in the corona, it will be the first ion species to 
resonate with these waves and can be preferentially accelerated during a flare.

\begin{figure}[h!]
\plotone{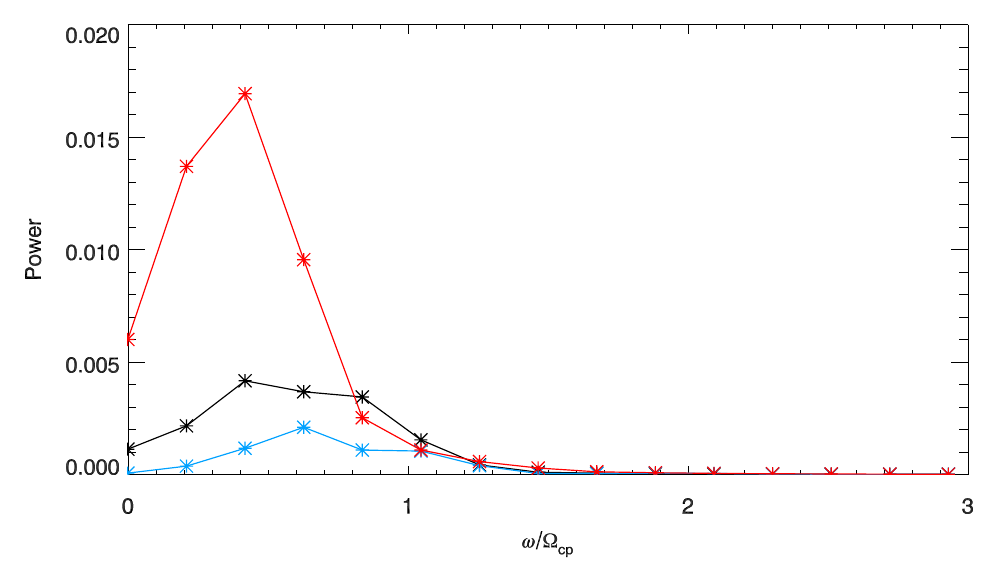}
\caption{The temporal FFT power spectrum at low frequencies for Run 1 at t = 50 to 80 
$\Omega_{cp}^{-1}$ (black), Run 2 at t = 60 to 90 $\Omega_{cp}^{-1}$ (blue), and Run 3 at t = 40 to 70 
$\Omega_{cp}^{-1}$ (red). FFTs were calculated in the same way as Fig.\ref{fig:omegakpar} and then 
summed over $k_x$.
\label{fig:freq}}
\end{figure}

As we have shown, however, ion beams that are produced during flare energy release are efficient 
drivers of oblique right- and left-handed waves in the range of the proton cyclotron frequency 
and below (see Fig.~\ref{fig:freq}). Thus, similar arguments suggest that the waves on the higher end 
of this frequency spectrum should drive $^3$He abundance enhancements, with waves in the frequency 
range $0.5\Omega_{cp} < \omega < 0.67\Omega_{cp}$ affecting particle populations above the flare site 
and waves in the frequency range $0.67 \Omega_{cp} < \omega < \Omega_{cp}$ affecting particle 
populations below the flare site. However, because these waves are oblique, both right- and left-handed 
waves can resonate with cold $^3$He. In the case of left-handed waves, it is the $n = +1$ resonance in 
Eq.~\ref{eq:resonance}, while for right-handed waves it is the $n = -1$ resonance that will heat cold 
$^3$He. The heated particles will then stream into the flare energy release region where they 
will be further accelerated to observed energies by Fermi reflection during reconnection 
\citep{drake_electron_2006,drake_power-law_2012}.

In our simulations, the $^3$He is heated to a thermal speed $\sim v_A$ (see Fig.~\ref{fig:h3vdist}
and Fig.~\ref{fig:h3temps}). Thus, the total energy flux of the heated particles moving into the flare 
region will be $(1/4)n_{3He}m_{3He}v_A^3$, with an additional factor of 1/2 accounting for some 
particle transport in the opposite direction. The electromagnetic energy of the waves responsible for 
heating is $\Tilde{B}^2/8\pi$, where $\Tilde{B}$ is the average magnetic field amplitude of waves with 
frequencies $0.5\Omega_{cp} \leq \omega \leq \Omega_{cp}$. Like Alfv\'en waves, the plasma energy of 
cyclotron waves is roughly equal to the field energy. Using $2v_A$ as the wave velocity, this gives a 
wave energy flux equal to $\Tilde{B}^2v_A/2\pi$. Equating this with the $^3$He energy flux and 
neglecting factors of order unity, we find that the number density of heated $^3$He is directly 
proportional to the squared amplitude of the wave magnetic field: 
$n_{3He}/n_p \sim \Tilde{B}^2/B_0^2$, where $n_p$ is the number density of the protons and $B_0$ is 
the background magnetic field. We emphasize that $\Tilde{B}$ in this expression only corresponds to the wave energy in the frequency band $0.5\Omega_{cp} \leq \omega \leq \Omega_{cp}$.

Using the FFT frequency spectra, we can obtain $\Tilde{B}^2$ by calculating the fraction of power in the 
frequency range $0.5\Omega_{cp} \leq \omega \leq \Omega_{cp}$. To do this, we fit 
the data presented in Fig.~\ref{fig:freq} from $\omega = 0$ to $\omega = 1.5\Omega_{cp}$ with 
polynomials. We then calculate the fraction of the energy in the range $0.5\Omega_{cp} \leq \omega \leq 
\Omega_{cp}$ using the polynomial representation. Excluding the lowest energy case, which did not 
produce good $^3$He heating, these fractions are 46\% for Run 1 and 31\% for Run 3. From 
Fig.~\ref{fig:anisotropy}d, the average value of $B_z^2/B_0^2$ over the time period of the FFTs is 8.5\% 
for Run 1 and 22\% for Run 3; therefore, $\Tilde{B}^2/B_0^2$ is 3.91\% for Run 1 and 6.82\% for Run 3. 
These percentages are on the order of the $^4$He density in the corona. We emphasize, of course, that in 
our picture of $^3$He abundance enhancement, the waves are responsible for driving $^3$He into the flare 
site and raising their energy to the range of $m_{3He}v_A^2\sim 10$ keV/nucleon, but it is 
reconnection-driven energization that further accelerates these particles to energies of the order of 0.1-10 MeV/nucleon \citep{mason_3he-rich_2007}.

Thus, energetic ion beams are a strong candidate for driving the waves responsible for the 
enhancements of $^3$He observed in impulsive flares. However, further simulations are needed to  
explore how the $^3$He is resonantly heated in an inhomogeneous magnetic field. In addition, a numerical 
analysis of the dispersion relation using Eq.~\ref{eq:distfunc} should be able to establish definitively the 
instabilities responsible for the waves observed in our simulations. Ongoing work in this area using 
the {\tt Arbitrary Linear Dispersion Solver (ALPS)} ~\citep{verscharen_alps_2018} will be covered in a 
future paper.


\section{Acknowledgments}
The authors were supported by NSF grant PHY2109083 and NASA grant 80NSSC20K1813 and NASA FINESST award 80NSSC23K1625. This study benefits
from discussions within the International Space Science
Institute (ISSI) Team ID 425 “Origins of 3He-rich SEPs.” We acknowledge informative discussions with Drs. R.\ Bucik and G.\ Mason. 
The PIC simulations were performed at the National Energy Research Supercomputing Center.

\bibliography{apjdraft}
\bibliographystyle{aasjournal}



\end{document}